\documentclass[10pt]{article}

 \usepackage{amsmath,amsthm,amssymb}
 \usepackage{makeidx,epsfig,lscape}
 \usepackage{color,colortbl}
 \usepackage{fancyhdr}
 \usepackage{xcolor,pict2e}
\usepackage{adjustbox}
\usepackage{balance}
\usepackage{graphicx}
\usepackage{changepage}
\usepackage{subcaption}
\usepackage{multicol}
\usepackage{floatrow}

\newfloatcommand{capbtabbox}{table}[][\FBwidth]

 \renewcommand{\headrulewidth}{0pt}
 \renewcommand{\footrulewidth}{0.5pt}

 \definecolor{myaqua}{rgb}{0.0,0.5,0.55}
 \definecolor{lightaqua}{rgb}{0.75,0.95,0.95}

 \usepackage[colorlinks = true,
            linkcolor = myaqua,
            urlcolor  = blue,
            citecolor = myaqua]{hyperref}

\usepackage{caption}
\usepackage{floatrow}
\def\lin#1#2{\textcolor[rgb]{0.6,0.6,0.6}{\vspace*{#1mm} \hrule
   height 3 pt \vspace*{#2mm}}}
\def\bt{\begin{tabular}}
\def\et{\end{tabular}}
\def\and{\mbox{ and }}

\def\1{{\bf 1}}

 \def\sectionn#1{\refstepcounter{section}{\color{myaqua}

 \vskip 6mm

 \noindent\Large\bf\thesection. #1}

 \vskip 3mm}

 \def\boxx#1#2#3#4#5{
 {\linethickness{#4pt}\put(#1,#5){\color{myaqua}{\line(1,0){#3}}}}
 \multiput(#1,#2)(0,#4){2}{\line(1,0){#3}}
 \multiput(#1,#2)(#3,0){2}{\line(0,1){#4}}
  }

\begin{document}

 $\mbox{ }$

 \vskip 12mm

{ 

{\noindent{\huge\bf\color{myaqua}
  The graph structure of the Internet at the Autonomous Systems level during ten years}}
%
\\[6mm]
{\large\bf Agostino Funel$^1$}}
\\[2mm]
{ 
 $^1$ENEA, Energy Technologies Department,
 ICT Division-HPC Lab, 
 Italy\\
Email: \href{mailto:agostino.funel@enea.it}{\color{blue}{\underline{\smash{agostino.funel@enea.it}}}}\\[1mm]
 \\[4mm]

 \includegraphics{pic2.ps}

\lin{5}{7}

 { 
 {\noindent{\large\bf\color{myaqua} Abstract}{\bf \\[3mm]
 \textup{We study how the graph structure of the Internet at the Autonomous Systems (AS) level evolved during a decade. For each year of the period 2008-2017 we consider a snapshot of the AS graph and examine how many features related to structure, connectivity and centrality changed over time. The analysis of these metrics provides topological and data traffic information and allows to clarify some assumptions about the models concerning the evolution of the Internet graph structure. We find that the size of the Internet roughly doubled. The overall trend of the average connectivity is an increase over time, while that of the shortest path length is a decrease over time. The internal core of the Internet is composed of a small fraction of big AS and is more stable and connected that external cores. A hierarchical organization emerges where a small fraction of big hubs are connected to many regions with high internal cohesiveness, poorly connected among them and containing AS with low and medium number of links. Centrality measurements indicate that the average number of shortest paths crossing an AS or containing a link between two of them decreased over time.
 }}}
\\[4mm]
 {\noindent{\large\bf\color{myaqua} Keywords}{\bf \\[3mm]
 Network analysis, Graph theory, Internet, Autonomous Systems.
}

\lin{5}{7}

\renewcommand{\headrulewidth}{0.5pt}
\renewcommand{\footrulewidth}{0pt}

 \pagestyle{fancy}
 \fancyfoot{}
 \fancyhead{} 
 \fancyhf{}
 \fancyhead[RO]{\leavevmode \put(-90,0){\color{myaqua}A. Funel} \boxx{15}{-10}{10}{50}{15} }
 \fancyhead[LE]{\leavevmode \put(0,0){\color{myaqua}A. Funel}  \boxx{-45}{-10}{10}{50}{15} }
 \fancyfoot[C]{\leavevmode
 \put(0,0){\color{lightaqua}\circle*{34}}
 \put(0,0){\color{myaqua}\circle{34}}
 \put(-2.5,-3){\color{myaqua}\thepage}}

 \renewcommand{\headrule}{\hbox to\headwidth{\color{myaqua}\leaders\hrule height \headrulewidth\hfill}}

\sectionn{Introduction}\label{sec:intro}
{ \fontfamily{times}\selectfont
 \noindent The Internet is a highly engineered communication infrastructure continuously growing over time.
It consists of Autonomous Systems (ASes)  each of which can be considered  a network, with its own routing policy, administrated by a single authority.  ASes peer with each other to exchange  traffic and use the Border Gateway Protocol (BGP)~\cite{bgp} to exchange routing and reachability information in the global routing system of the Internet. Therefore, the Internet can be represented by a graph where ASes are nodes and BPG peering relationships are links.
The structure of the Internet has been studied by many authors and the literature on the subject is vast. One of the most used methods is the statistical analysis  of different metrics characterizing the  AS graph~\cite{Vazquez_et_al_2002},~\cite{Zhang_et_al_2005},~\cite{Mahadevan_2006},~\cite{Magoni_et_al_2001}. There are not many studies concerning the evolution of the Internet over time~\cite{Edwards_et_al_2012},~\cite{Dhamdhere_2008},\cite{Guo_et_al_2008} and because the amount of data to analyze tends to grow dramatically often only a limited number of properties are considered. The purpose of this work is to study the evolution of the Internet considering features related to both its topology and data traffic.
To achieve this goal we consider for each year of the period 2008-2017 a snapshot of the undirected AS graph, introduce three classes of metrics related to structure, connectivity, centrality and analyze how they change over time.
The paper is organized as follows: in Section~\ref{sec:datasets} we describe the datasets; in Section~\ref{sec:metrics} we define the adopted metrics and for each of them explain its importance; we report the results in Section~\ref{sec:results}. Finally, in Section~\ref{sec:conclusion} we summarize the results and make the final considerations.}

\section{Data sets}\label{sec:datasets}
The ASes graphs have been constructed from the publicly available  IPv4 Routed /24 AS Links Dataset provided by CAIDA~\cite{caida}. AS links are derived from traceroute-like IP measurements collected by the Archipelago (Ark)~\cite{ark} infrastructure, a globally distributed hardware platform of network path probing monitors. The association of an IP address with an AS is based on the RouteViews~\cite{route_view} BGP data and the probed IP paths are mapped into AS links. We exclude multi-origin ASes and AS sets because they may introduce distortion in the association process due to the fact that the same prefix could be advertised by many different ASes creating an ambiguity in the mapping process between IP addresses and ASes.  The sizes of the ASes graphs analyzed in this work are shown in Tab.~\ref{tab_datasets}.

\begin{table*}[htbp]
\begin{center}
\begin{adjustbox}{max width=\textwidth}
\begin{tabular}{|c||c|c|c|c|c|c|c|c|c|c|}
\hline
Year            & 2008   & 2009  & 2010   & 2011   & 2012    & 2013     & 2014  & 2015  & 2016  & 2017 \\
\hline
\# Nodes        & 28838  & 31892 & 35149  & 38550  & 41527   & 47407    & 47581 & 50856 & 51736 & 52361 \\
\hline
\# Edges        & 135723 & 152447 & 184071 & 213870 & 281596 & 282939 & 298355 & 347518 & 379652 & 414501 \\    
\hline
\end{tabular}
\end{adjustbox}
\end{center}
\caption{Sizes of the ASes undirected graphs.}
\label{tab_datasets}
\end{table*}

\section{Description of metrics}\label{sec:metrics}
In this section we introduce the metrics chosen for this analysis whose summary scheme is shown in Tab.~\ref{tab_metrics}. For each metric we give a short description and briefly discuss its importance.
We use the notation $G=(N,E)$ to indicate an AS graph which has $N$ nodes and $E$ edges.

\begin{table*}[h!]
\begin{center}
\begin{adjustbox}{max width=\textwidth}
\begin{tabular}{|c|c|c|}
\hline
Metric & Relevance & Importance \\
\hline
\hline
Degree distribution  & Structure  & Scale-free, global properties. \\
k-core decomposition &  & Nested hierarchical structure of tightly interlinked subgraphs.\\
\hline
Clustering coefficient & Connectivity & Neighbourhood connectivity. Hierarchical structure. \\
Shortest path length    &              & Reachability (minimum number of hops between two ASes). \\
\hline
Closeness centrality  & Centrality & Indicates the proximity of an AS to all others.\\
Node betweenness centrality &      & Related to node traffic load. \\
Edge betweenness centrality &      & Related to link traffic load. \\
\hline
\end{tabular}
\end{adjustbox}
\end{center}
\caption{Metrics used to study the evolution of the Internet at the AS level over time.}
\label{tab_metrics}
\end{table*}

\subsection{Degree distribution}\label{subsec:degree_distr}
The degree distribution $P(k)$ is the probability that a random chosen node has degree $k$. If a graph has $N_k$ nodes with degree $k$ then $P(k)=N_k/N$. Since $P(k)$ is a probability distribution it satisfies the normalization condition $\sum_{k=k_{min}}^{k_{max}} P(k) = 1$ where $k_{min}$ and $k_{max}$ are the minimum and maximum degree, respectively. From $P(k)$ we can calculate the average degree $\hat{k}=\sum_{k=k_{min}}^{k_{max}} k P(k)$. For a random network $P(k)$ follows a binomial distribution and in the limit of sparse network $\delta \ll 1$, where $\delta$ is the link density, it is well approximated by a Poissonian. The Internet, as many other real networks, can be considered sparse and, moreover, it is \emph{scale-free} which means that it contains both small  and very high degree nodes and this feature can not be reproduced by a Poissonian. Many studies agree that the degree distribution follows a power law $P(k) \propto k^{-\alpha}$ though deviations have been observed~\cite{Faloutsos_1999},~\cite{Chen_2002},~\cite{Mahadevan_2006}. For each snapshot of the AS graph we calculate the best fit power law parameters $k_{min}^{PL}$ and $\alpha$  and  verify the statistical plausibility of this model. 

\subsection{K-core decomposition}\label{subsec:kcore}
A k-core of a graph is obtained by removing all nodes with degree less than $k$. Therefore, the k-core is the maximal subgraph in which all nodes have at least degree $k$. The 0-core is the full graph and coincides with the 1-core if there are no isolated nodes, as in the case of the Internet. The k-core decomposition is a way of \emph{peeling} the graph by progressively removing the outermost low degree layers up to the innermost high degree core which we call nucleus. We denote by $k_{max}^{core}$ the coreness of the nucleus, and by $\mathcal{N}_n$ ($\mathcal{N}_k$) and $\mathcal{E}_n$ ($\mathcal{E}_k$) the number of ASes and edges in the nucleus (in the k-core). In the case of the Internet the analysis of the k-core decomposition over time is useful for understanding whether its nucleus, composed of high degree ASes, evolves differently from its periphery.  

\subsection{Clustering coefficient}\label{subsec:clustcoeff}
The local clustering coefficient $C_i$ of a node $i$ of degree $k$ is the ratio of the actual number of edges $E_i$ connecting its neighbours to the maximum possible number of edges that could connect them. For an undircted graph $C_i = 2 E_i/k (k -1)$. By averaging over all nodes we obtain the global clustering coefficient  $C = \sum_i C_i / N$. For a random network $C$ is independent of the node's degree and decreases with the size of the graph as $C \sim N^{-1}$. Scale-free networks exhibit a quite different behavior. For example the clustering coefficient of  a scale-free network obtained from the Barabasi-Albert model~\cite{Barabasi_Albert_99} follows $C \sim (\mbox{ln} N)^2/N$, which for large $N$ is higher than that of a random network. An important quantity is $C(k)$, the average clustering coefficient of degree $k$ nodes. It has been shown~\cite{Vazquez_etal2_2002} that it is the three-points correlation function which is the probability that a degree $k$ node is connected to two other nodes which in their turn are joined by an edge. $C(k)$ can be used to study the hierarchical structure of networks~\cite{Vazquez_et_al_2002},~\cite{Ravasz_etal2003}.  

\subsection{Shortest path length}\label{subsec:spl}
The shortest path length between two nodes is the minimum number of hops needed to connect them. Of course, for any pair of nodes there may be several shortest paths connecting them. The shortest path length distribution $s(h)$ provides, for a given number $h$ of hops, the number of shortest paths of length $h$. We call $S$ the average shortest path length. The diameter $D$ is the longest shortest path. The importance of the shortest paths is mainly related to routing. Many routing algorithms are based on the shortest path length. Adaptive algorithms allow to change routing decision to optimize traffic load and prevent incidences of congestions. The knowledge of the  available shortest paths is then crucial for  routing efficiency.

\subsection{Closeness centrality}\label{subsec:closeness_centr}
The closeness centrality $\Gamma$ of a node $i$ is the inverse of its average shortest path length to all other nodes: $\Gamma(i) = (N-1)/ \sum_{j=1}^{N-1} \sigma(i,j)$ where $\sigma(i,j)$ is the shortest path length between $i$ and $j$. Nodes with high $\Gamma$ are those closest to all others and can be considered central in the network. On the contrary, nodes with low $\Gamma$ are, on average, far away from the others and can be considered peripheric. 

\subsection{Betweenness centrality}\label{subsec:betw_centr}
The concept of betweenness centrality applies to both nodes and edges. The betweenness centrality of a node $i$ is defined as $B_n(i) = \sum_{j,k \in V} \sigma(j,k \mid i) / \sigma(j,k)$ where the sum is over all pairs of nodes, $\sigma(j,k)$ is the number of shortest paths and $\sigma(j,k \mid i)$ is the number of those passing through $i$. If $j=k$ then $\sigma(j,k)=1$ and if $i \in \{j,k\}$, $\sigma(j,k \mid i)=0$. The betweenness centrality $B_e$ of an edge $e$ is defined in the same way. In this case $\sigma(j,k \mid e)$ is the number of shortest paths containing $e$.
Efficient routing policies exploit as much as possible available shortest paths, hence a node (edge) with high betweenness centrality carries large traffic load. In~\cite{Holme_2002} the betweenness centrality was used to investigate the evolution of networks whose nodes may break down due to overload and in~\cite{Goh_2001} it was used to  define the load of a node for studying the problem of data packet transport in power law scale free networks.

\section{Results}\label{sec:results}
In this section we compare the measurements of the metrics concerning the Internet AS graphs obtained for each year of the decade 2008-2017 and report the corresponding results.
 
\subsection{Degree distribution}\label{subsec:res_degree_distr}
Fig.~\ref{fig_deg_and_ccdf_distr} shows the node probability degree distributions and their complementary cumulative functions (CCDF). For all data sets the peak of the degree distribution is for $k=2$, a result already reported in~\cite{Mahadevan_2006} where it is claimed that it is due to the AS number assignment policies. While the edge density is around $\delta \approx 3 \times 10^{-4}$ during the decade 2008-2017, the general trend is a growth over time for both $\hat{k}$ and $k_{max}$ as shown in Tab.~\ref{tab_degree}. This means that the Internet has become more connected preserving its sparse nature. All degree distributions have a similar form. In order to verify the statistical plausibility of the power law model we perform a goodness of fit test based on the Kolmogorov-Smirnov statistic~\cite{Clauset2009} which provides a  $p$-value. The power law has statistical support if $p > 0.1$. From Tab.~\ref{tab_degree} we see that even if the best fit exponent  is always around the value $\alpha \approx 2.1$ the power law can be considered a reliable model only for the distributions of the years 2008, 2010 and 2011. Since for the majority of the most larger data sets $ p \leq 0.1$  we could say that at the AS level the evolution of the Internet can not be explained by models which predict a pure power law degree distribution.

\begin{figure}[htbp]
  \begin{center}
    \includegraphics[width=3.5in]{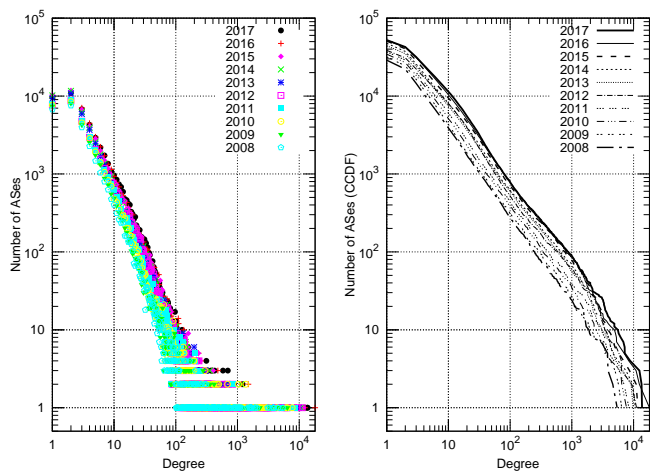}
\caption{Node degree probability distributions (left) and their  CCDF (right).}
\label{fig_deg_and_ccdf_distr}
  \end{center}
\end{figure}

\begin{table*}[h!]
\begin{center}
\begin{adjustbox}{max width=\textwidth}
\begin{tabular}{|c||c|c|c|c|c|c|c|c|c|c|}
\hline
Year            & 2008   & 2009  & 2010   & 2011   & 2012    & 2013     & 2014  & 2015  & 2016  & 2017 \\
\hline
$\delta\times10^{-4}$ & 3.3 & 3.0 & 3.0 & 2.9 & 3.3 & 2.5 & 2.6 & 2.7 & 2.8 & 3.0 \\
\hline
$\hat{k}$        & 9.4  & 9.6 & 10.5  & 11.1  & 13.6   & 11.9    & 12.5 & 13.7 & 14.7 & 15.8 \\
\hline
$k_{max}$         & 5430 & 6430 & 7684 & 8416 & 11179 & 9838 & 10682 & 11739 & 18110 & 13725 \\
\hline
$k_{min}^{PL}$         & 23 & 8 & 44 & 30 & 16 & 15 & 17 & 17 & 14 & 14 \\
\hline
$\alpha$         & $2.12\pm0.03$ & $2.13\pm0.01$ & $2.08\pm0.03$ & $2.07\pm0.03$ & $2.20\pm0.02$ & $2.10\pm0.01$ & $2.10\pm0.02$ & $2.11\pm0.01$ & $2.10\pm0.01$ & $2.14\pm0.01$ \\
\hline
$p\pm0.01$         & $0.67$ & $0.04$ & $0.60$ & $0.93$ & $0$ & $0$ & $0$ & $0$ & $0$ & $0$ \\
\hline
\end{tabular}
\end{adjustbox}
\end{center}
\caption{The table shows: the edge density $\delta$, the average degree $\hat{k}$, the maximum degree $k_{max}$. The best fit power law cut off and exponent are $k_{min}^{PL}$ and $\alpha$. The condition $p > 0.1$ indicates statistical plausibility of the power law model.}
\label{tab_degree}
\end{table*}

\subsection{K-core decomposition}\label{subsec:res_kcore}
The left plot of Fig.~\ref{fig_kcore_orders_nodes_edge_distr_and_kcore_nucleus} shows for each year of the decade 2008-2017 the distributions of ASes and edges in each k-core. We observe that in general for each k-core both the number of ASes and edges increase over time. The evolution of the Internet nucleus is shown in the right plot of Fig.~\ref{fig_kcore_orders_nodes_edge_distr_and_kcore_nucleus}. The coreness of the nucleus   increases over time (in 2016 and 2017 it has the same value). The fraction of ASes in the nucleus is quite stable over time although in absolute value $\mathcal{N}_n$ increases from 2008 to 2013, decreases in 2014 and 2015 and then increases again until 2017. We observe the same trend also for the number of edges in the nucleus as shown in Tab.~\ref{tab_nucleus_and_perc_var_kcore}.

\begin{figure}
\centering
\begin{subfigure}{.5\textwidth}
  \centering
  \includegraphics[width=3in]{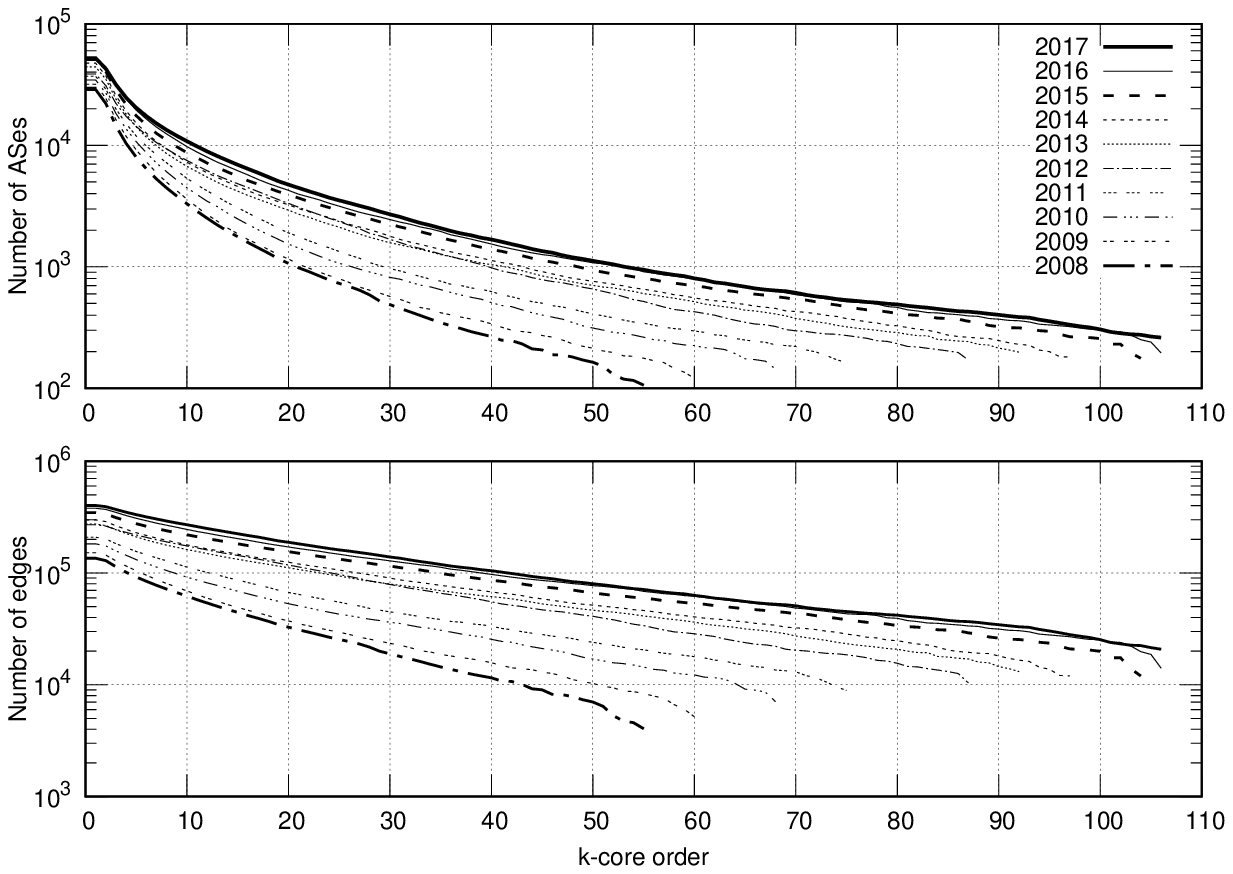}
\end{subfigure}%
\begin{subfigure}{.5\textwidth}
  \centering
\includegraphics[width=3in]{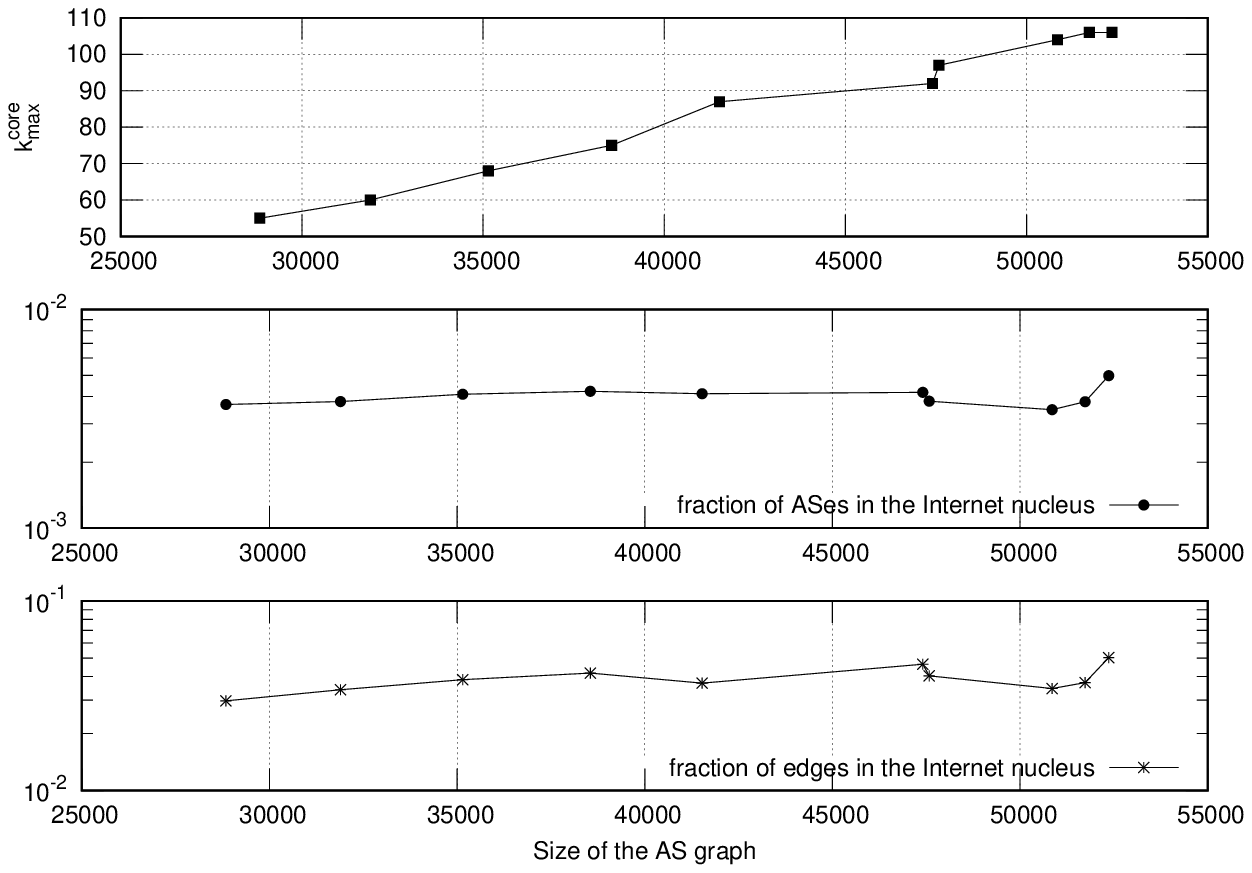}
\end{subfigure}
\caption{\textbf{Left}: number of ASes and edges in each k-core during the decade 2008-2017. \textbf{Right}: for each year of the decade 2008-2017 are shown, as a function of the size of the AS graph: the highest coreness  $k_{max}^{core}$ (top); the percentage of ASes (middle) and edges (bottom) in the Internet nucleus.}
\label{fig_kcore_orders_nodes_edge_distr_and_kcore_nucleus}
\end{figure}

On average the nucleus contains $\sim$0.4\% of all ASes and $\sim$4\% of all edges. Carmi {\itshape et al.}~\cite{Carmi_2007} predicted the increase of $k_{max}^{core}$ and $\mathcal{N}_n$ as a power of $N$ on the base of a numerical simulation assuming a scale-free growing model with the same parameters of the real Internet. Instead, from the analysis of the Internet at the AS level during the period 2001-2006 Guo-Quing Zhang {\itshape et al.}~\cite{Guo_et_al_2008} found no clear evidence of the exponential growth of $k_{max}^{core}$ and observed a stability of its value after 2003. They also found that the size of the nucleus exhibits large fluctuations over time.
We now examine in the left plot of Fig.~\ref{fig_kcore_periphery_all_and_kcore_edgedensity_all} how the fraction of ASes and edges varies in the periphery of the Internet from  the $2$ to the $10$-core. We start from the core of order 2 because in our case there are no isolated ASes.  Compared to the evolution of the nucleus it is evident that the periphery evolves with a different dynamics.  In Tab.~\ref{tab_nucleus_and_perc_var_kcore} we compare for each k-core the number of ASes and edges it contained in 2008 and 2017 and report the percentage variation. Results clearly show that the nucleus is much more stable than the periphery. The connectivity of each core can be studied by looking at its edge density which is defined as $\delta_k = 2 \mathcal{E}_k/ \mathcal{N}_k (\mathcal{N}_k - 1)$. In the right plot of Fig.\ref{fig_kcore_periphery_all_and_kcore_edgedensity_all} is shown $\delta_k$ as a function of $N$ and in Tab.~\ref{tab_nucleus_and_perc_var_kcore} is reported its average value. The edge density increases with the coreness showing that the inner is the core the more it is connected. It is interesting to note that the edge density of the Internet nucleus is three order of magnitude higher than that of the most external 2-core. From a topological point of view this might imply the existence of an underlying hierarchical organization of the Internet with a small fraction of big ASes tightly connected among them and many regions composed of ASes with low or medium number of links. This structural property is investigated in more detail in the next section.

Tauro {\itshape et al.}~\cite{Tauro_et_al_2001} studied the topology of the Internet from the end of 1997 to the middle of 2000. They introduced the concept of importance of a node on the base of its degree and effective eccentricity defined as the minimum number of hops required to reach at least 90\% of all other nodes. The most important nodes have high degree and low effective eccentricity. The found that the  structure of the Internet is hierarchical with a highly connected core surrounded by layers of nodes of decreasing importance.

\begin{figure}
\centering
\begin{subfigure}{.5\textwidth}
  \centering
\includegraphics[width=3in]{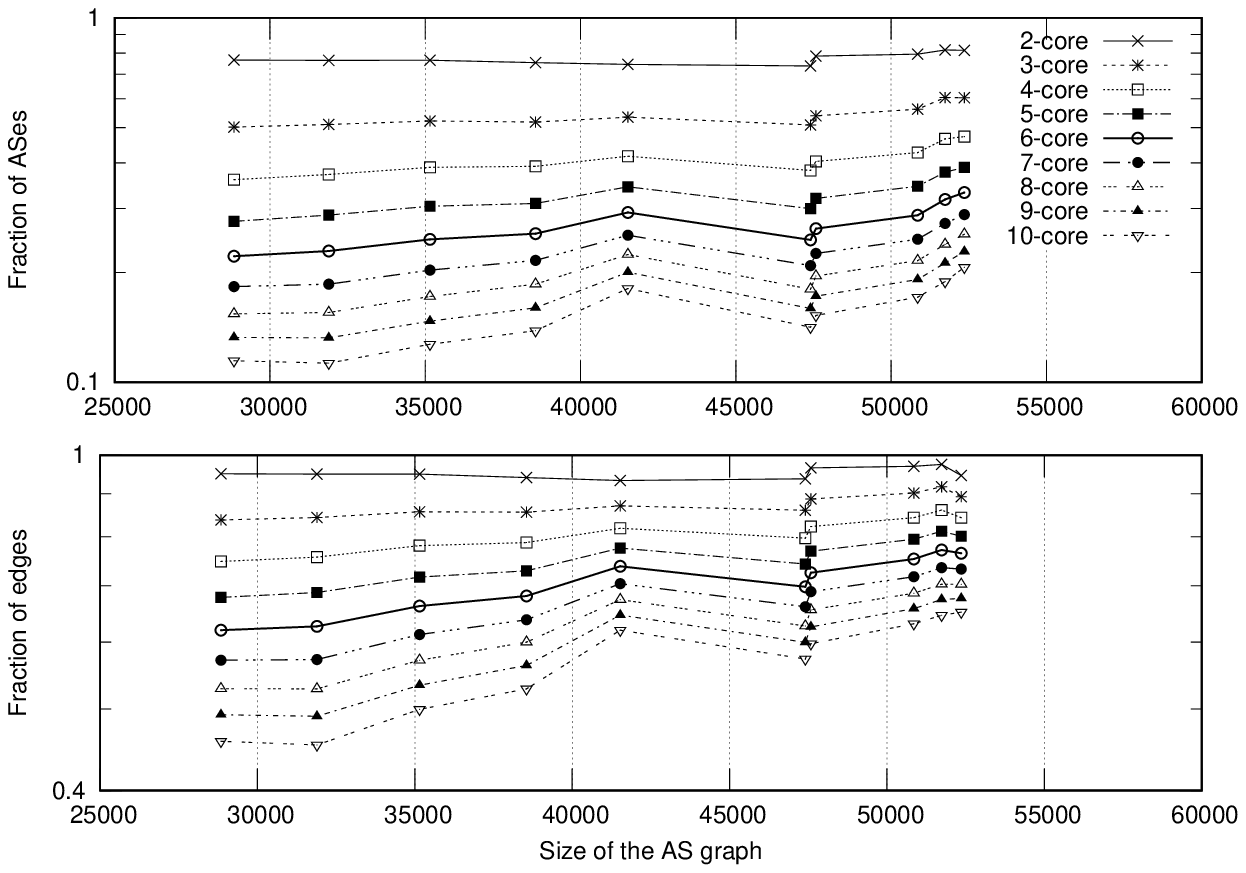}
\end{subfigure}%
\begin{subfigure}{.5\textwidth}
  \centering
\includegraphics[width=3in]{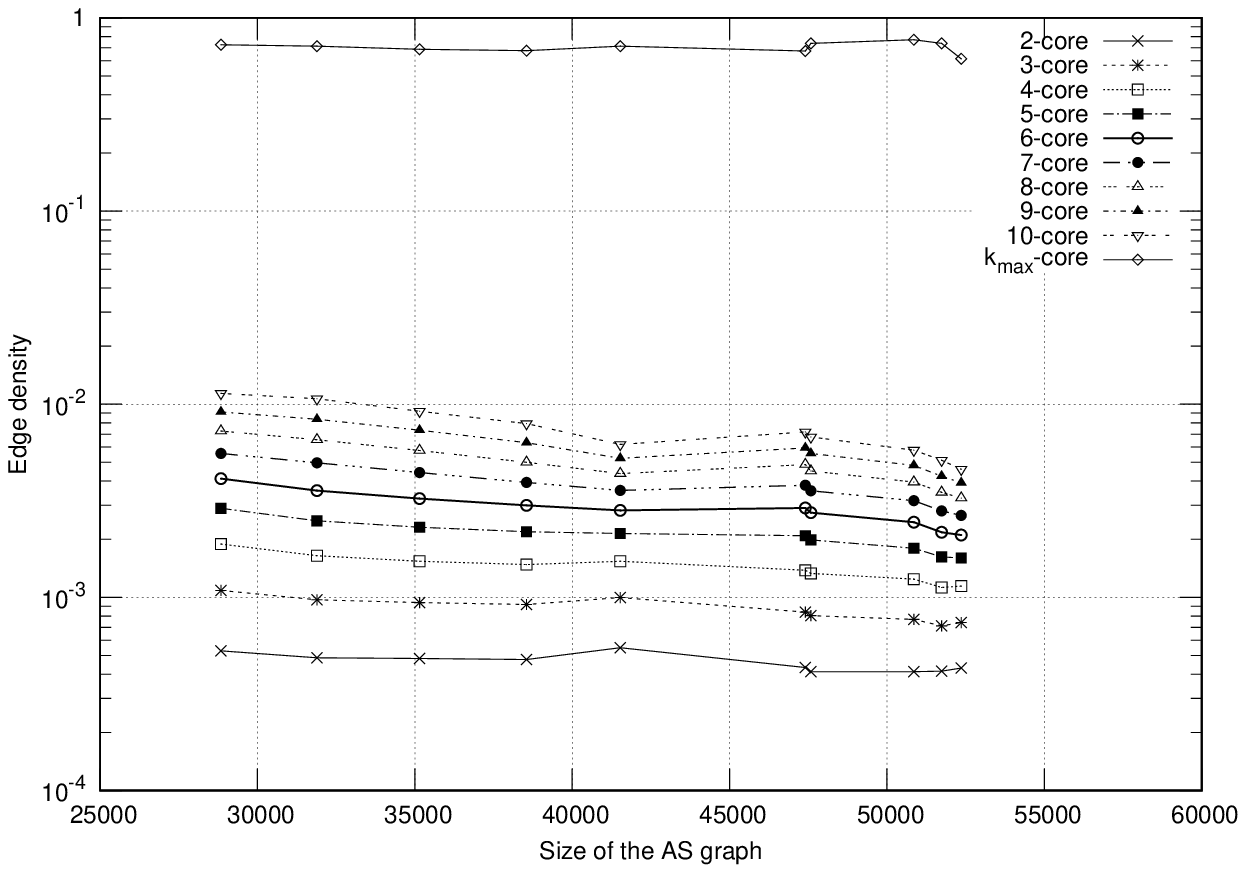}
\end{subfigure}
\caption{\textbf{Left}: fraction of ASes and edges in the periphery of the Internet from the 2 to the 10-core. \textbf{Right}: edge density $\delta_k$ as a function of the size of the AS graph for the k-cores and the Internet nucleus.}
\label{fig_kcore_periphery_all_and_kcore_edgedensity_all}
\end{figure}

\begin{table}[htbp]
\begin{center}
\begin{adjustbox}{max width=\textwidth}
\begin{tabular}{|c|c|c|c|||c|c|c|c|}
\hline
\rowcolor{lightaqua}
Year            &  $k_{max}^{core}$  & $\mathcal{N}_n$  & $\mathcal{E}_n$  &  $k$  (core)          &  $\Delta \mathcal{N}_k$(\%)  & $\Delta \mathcal{E}_k$(\%) & $\delta_k$ \\
\hline
\hline
2008            & 55  & 106 & 4040  & 2 & 4.82 & -0.4 &  $(42.1 \pm 4.7)\times 10^{-5}$ \\
\hline
2009 & 60 & 121 & 5185 & 3 & 10.22 & 5.48 & $(79.8 \pm 9.2)\times 10^{-5}$ \\
\hline
2010 & 68 & 144 & 7077 & 4 & 11.24 & 9.49 & $(13.0 \pm 1.5)\times 10^{-4}$ \\
\hline
2011 & 75 & 163 & 8921 & 5 & 11.26 & 12.31 & $(19.2 \pm 2.3)\times 10^{-4}$ \\
\hline
2012 & 87 & 171 & 10383 & 6 & 11.00 & 14.49 & $(26.5 \pm 3.3)\times 10^{-4}$ \\
\hline
2013 & 92 & 198 & 13132 & 7 & 10.56 & 16.13 & $(35.0 \pm 4.6)\times 10^{-4}$ \\
\hline
2014 & 97 & 181 & 12020 & 8 & 10.08 & 17.44 & $(44.6 \pm 6.1)\times 10^{-4}$ \\
\hline
2015 & 104 & 177 & 12008 & 9 & 9.57 & 19.3 & $(55.3 \pm 7.8)\times 10^{-4}$ \\
\hline
2016 & 106 & 196 & 14084 & 10 & 9.22 & 19.38 & $(6.8 \pm 1.0)\times 10^{-3}$ \\
\hline
2017 & 106 & 261 & 20838 & $k_{max}^{core}$ & 0.13 & 2.05 & $(64.1 \pm 6.8)\times 10^{-2}$ \\
\hline
\end{tabular}
\end{adjustbox}
\end{center}
\caption{\textbf{Left}: for each year in the decade 2008-2017 are shown the coreness $k_{max}^{core}$ of the Internet nucleus, and the number of ASes $\mathcal{N}_n$ and  edges $\mathcal{E}_n$ it contains. \textbf{Right}: percentage variation in the number of ASes $\Delta \mathcal{N}_k$ and edges $\Delta \mathcal{E}_k$ in the core of order $k$ obtained comparing 2008 and 2017 data. The last column reports the average edge density $\delta_k$ calculated over all the years 2008-2017.}
\label{tab_nucleus_and_perc_var_kcore}
\end{table}

\subsection{Clustering coefficient}\label{subsec:res_clustcoeff}
The clustering coefficient has been used to investigate the hierarchical organization of real networks. The hierarchy could be a consequence of the particular role of the nodes in the network.  A stub AS does not carry traffic outside itself and is connected to a transit AS that, on the contrary,  is designed for this purpose. The hierarchy of the Internet is rooted in its geographical organization in international, national backbones, regional and local areas. This is the skeleton of the Internet. International and national backbones connected to regional networks which finally connect local areas to the Internet, implementing in such a way a best and less expensive strategy. It is reasonable to suppose that this hierarchical structure introduces correlations in the connectivity of the ASes.  A. V\'{a}zquez {\itshape et al.}~\cite{Vazquez_et_al_2002} showed that the hierarchical structure of the Internet is captured by the scaling  $C(k) \sim k^{- \gamma}$ and found $\gamma = 0.75$. Ravasz and Barabasi~\cite{Ravasz_etal2003} proposed a deterministic hierarchical model for which $C(k) \sim k^{-1}$ and using a stochastic version of the model showed that the hierarchical topology is again well described by the scaling $C(k) \sim k^{- \gamma}$ even if the value of $\gamma$ can be tuned by varying other network parameters.

\begin{figure}
\centering
\begin{subfigure}{.5\textwidth}
  \centering
\includegraphics[width=3in]{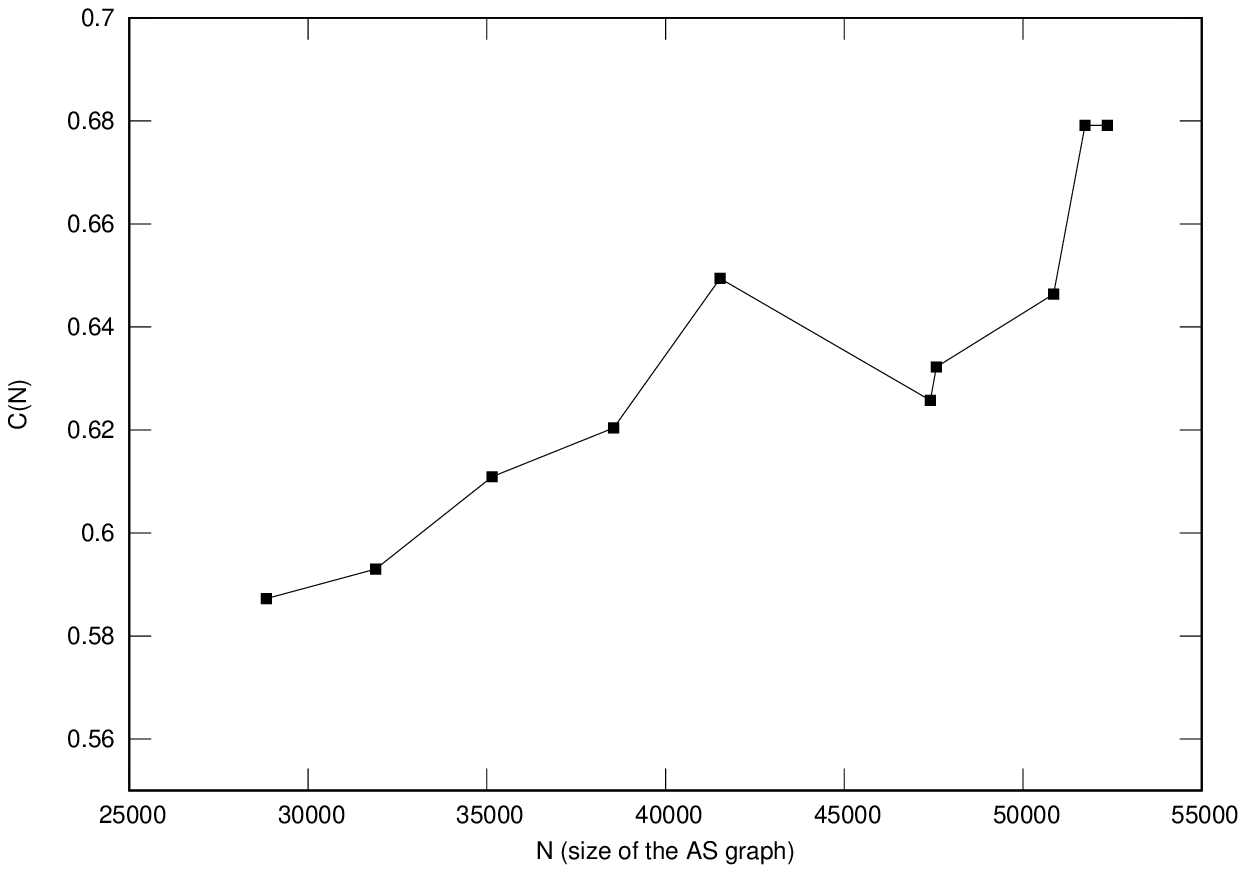}
\end{subfigure}%
\begin{subfigure}{.5\textwidth}
  \centering
\includegraphics[width=3in]{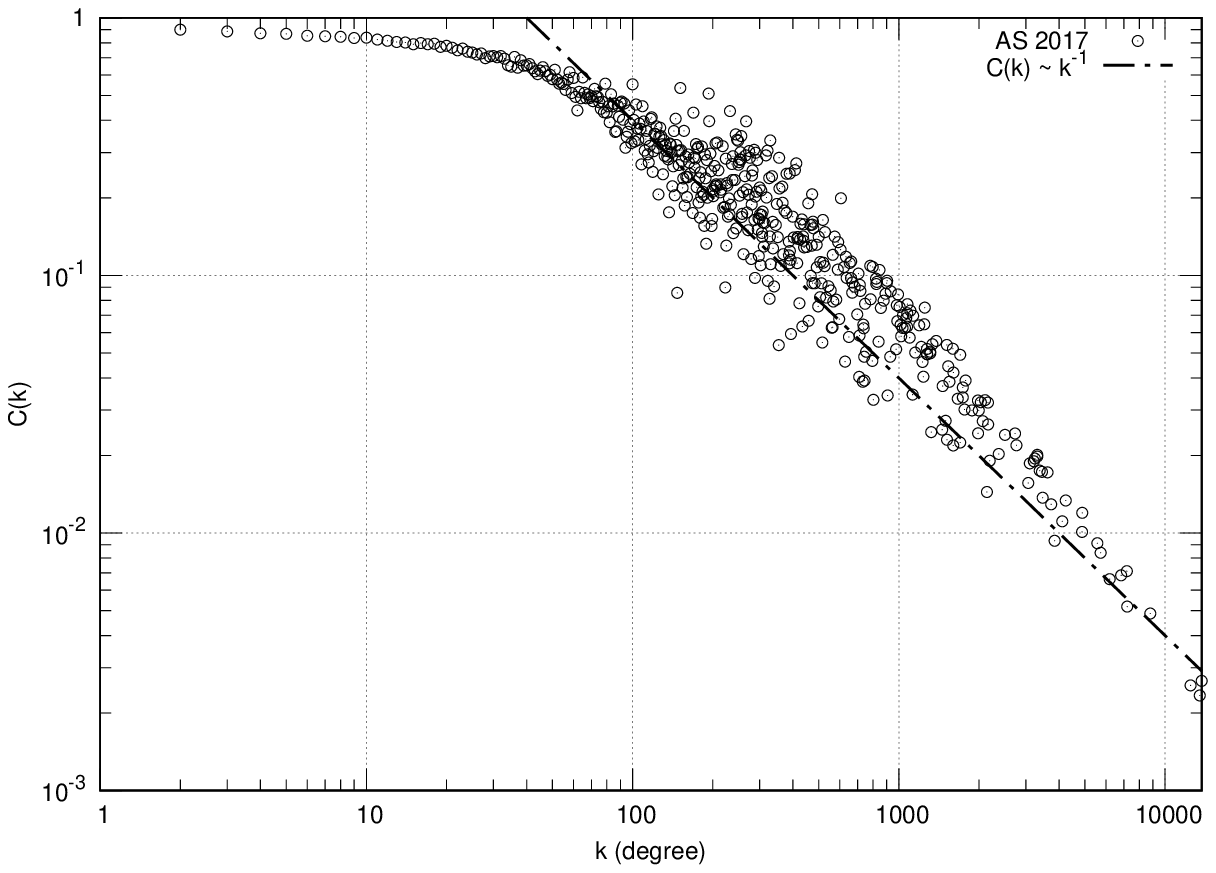}
\end{subfigure}
\caption{\textbf{Left}: average clustering coefficient as a function of the size of the AS graph during the decade 2008-2017. \textbf{Right}: average clustering coefficient as a function of the node's degree for the AS graph of the year 2017. The solid line shows the slope $C(k) \sim k^{-1}$.}
\label{fig_avg_clustcf_vs_n_and_avg_clustcf_vs_deg_single_fit}
\end{figure}

\begin{figure}[htpb]
  \begin{center}
    \includegraphics[width=4in]{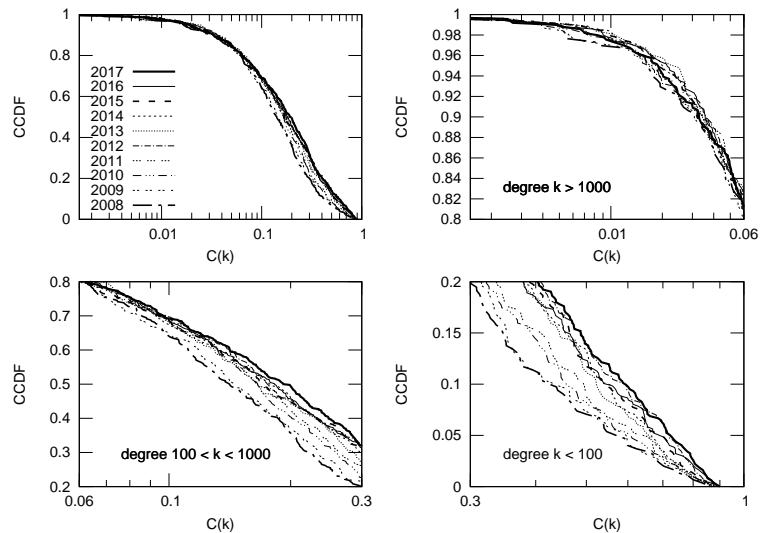}
\caption{CCDF of $C(k)$ for the years 2008-2017 (top left). The figure shows in more detail the high (top right), medium (bottom left) and low (bottom right) degree regions.}
\label{fig_avgclustcf_ccdistr_detail_withregions}
  \end{center}
\end{figure}

In the left plot of Fig.~\ref{fig_avg_clustcf_vs_n_and_avg_clustcf_vs_deg_single_fit} is shown the average clustering coefficient as a function of the size of the AS graph and in the fourth column of Tab.~\ref{tab_avgspl_avgclustcoeff} is reported its value for all the years 2008-2017. Apart from the year 2013, $C(N)$ weakly increases over time and the minimum and maximum values are $\sim$0.59 and $\sim$0.68 measured in 2008 and 2017, respectively. For the deterministic hierarchical model studied in~\cite{Ravasz_etal2003} $C$ is independent of $N$. The weak dependence of $C$  on $N$ might indicate the presence of a hierarchical organization in the structure of the Internet. To further investigate on this point we study $C(k)$. The right plot of Fig.~\ref{fig_avg_clustcf_vs_n_and_avg_clustcf_vs_deg_single_fit} shows $C(k)$ for the AS graph only for the year 2017 because for all other years the plots are almost overlapping. The best fit with the power law $k^{-\gamma}$ provides for all the years values of $\gamma$ which differ only by $\sim$0.1\% obtaining, on average, $\gamma = 1.08 \pm 0.01$. In the same figure is also shown the slope of the function $C(k) \sim k^{-1}$ and even if it nicely follows the slope of the experimental points the goodness of fit test does not give any statistical support to the scaling $C(k) = k^{-\gamma}$. However, data show that $C(k)$ decreases with $k$ especially for $k > 100$. Low degree ASes have high neighbourhood connectivity and, on the contrary, neighbours of big hub ASes are slightly connected among them. This is consistent with a hierarchical organization in which big ASes are connected to many regions with high internal cohesiveness and composed of low or medium degree ASes, and these regions are poorly connected among them. Since the $C(k)$ plots of the AS graph snapshots overlap, to study the evolution of the clustering coefficient over years we compare the CCDF of $C(k)$ in Fig.~\ref{fig_avgclustcf_ccdistr_detail_withregions}. For our convenience we consider in more detail three degree regions: high ($k > 1000$), medium $ 100 < k \leq 1000$, low $k \leq 100$ and also plot them in the same figure. We observe that in the high degree region the CCDF distributions are very intertwined indicating that during the decade 2008-2017 this region was rather static. In the medium degree region a clear separation emerges between the CCDF of the different years and for a given value of $C(k)$ the CCDF increases over time. The gap is even more pronounced in the peripheric low degree region. This result suggests that the evolution of the Internet  from 2008 to 2017 was not uniform and the most significant changes mainly affected its middle and even more its periphery, and the neighbourhood connectivity in these regions increased over time. 

\subsection{Shortest path length}\label{subsec:res_spl}
The left plot of Fig.~\ref{fig_shortest_paths_distr_and_hops_vs_n} shows the shortest path length distributions $s(h)$ for all the years 2008-2017 and in Tab.~\ref{tab_avgspl_avgclustcoeff} are reported the average values $S$ and the diameter $D$. 

\begin{table}[htbp]
\begin{center}
\begin{tabular}{|c|c|c|c|}
\hline
Year            &  $S \pm 0.6 $  & $D$ & $C$  \\
\hline
\hline
2008            & $3.1$  & 6  & 0.59  \\
\hline
2009 & $3.0$ & 7 & 0.59 \\
\hline
2010 & $3.0$ & 7 & 0.61 \\
\hline
2011 & $3.0$ & 6 & 0.62  \\
\hline
2012 & $2.9$ & 7 & 0.65  \\
\hline
2013 & $3.0$ & 6 & 0.63 \\
\hline
2014 & $3.0$ & 6 & 0.63  \\
\hline
2015 & $3.0$ & 6 & 0.65  \\
\hline
2016 & $2.9$ & 7 & 0.68  \\
\hline
2017 & $2.9$ & 7 & 0.68  \\
\hline
\end{tabular}
\end{center}
\caption{For each year in the decade 2008-2017 are shown the average shortest path length $S$, the diameter $D$ and the average clustering coefficient $C$.}
\label{tab_avgspl_avgclustcoeff}
\end{table}

We observe that the overall trend is a slight decrease of $S$ over time with an average of $\sim$3.0. Zhao {\itshape et al.}~\cite{Zhao_et_al_2008} analyzed BGP data from the Route-Views Project~\cite{route_view} in the period 2001-2006 and observed a very weak decreasing of $S$. They measured a decreasing rate of $\sim$2.5 $\times 10^{-4}$ and found $S(2001)\!=\!3.4611$ and $S(2006)\!=\!3.3352$. They noticed  that simple power law and small world models, which predict  a growth of $S$ with the size of the Internet, fail to explain the overall slight decrease of $S$ over time and argued that this might be due to the fact that the Internet expands according to many factors not considered by simple models like competitive and cooperative processes (like commercial relationships), policy-driven strategy and other human choices. From the comparison of our result with that of Zhao {\itshape et al.} there are indications the $S$ has been slightly reduced  during the period 2001-2017. This reinforces the fact that a pure power law model could not explain the evolution of the Internet because for  $2 < \alpha < 3$ it predicts $S \sim \mbox{ln}\mbox{ln}N$~\cite{Cohen_2003},~\cite{Bollobas_2004}. The right plot of Fig.~\ref{fig_shortest_paths_distr_and_hops_vs_n} shows, for different lengths, the number of shortest paths over time. The 3-hops shortest paths are the most numerous, as expected, and their number increases over time. 

\begin{figure}[htbp]
\centering
\begin{subfigure}{.5\textwidth}
  \centering
\includegraphics[width=3in]{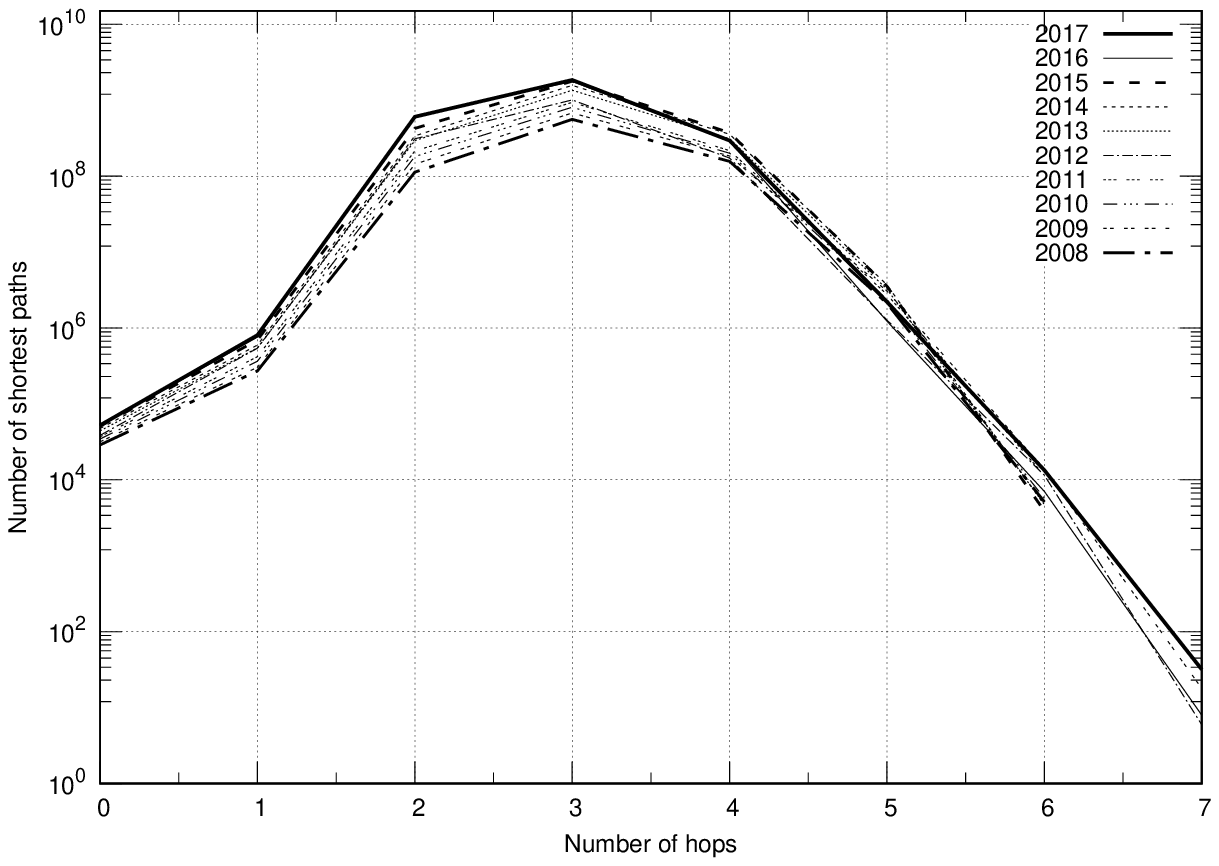}
\end{subfigure}%
\begin{subfigure}{.5\textwidth}
  \centering
\includegraphics[width=3in]{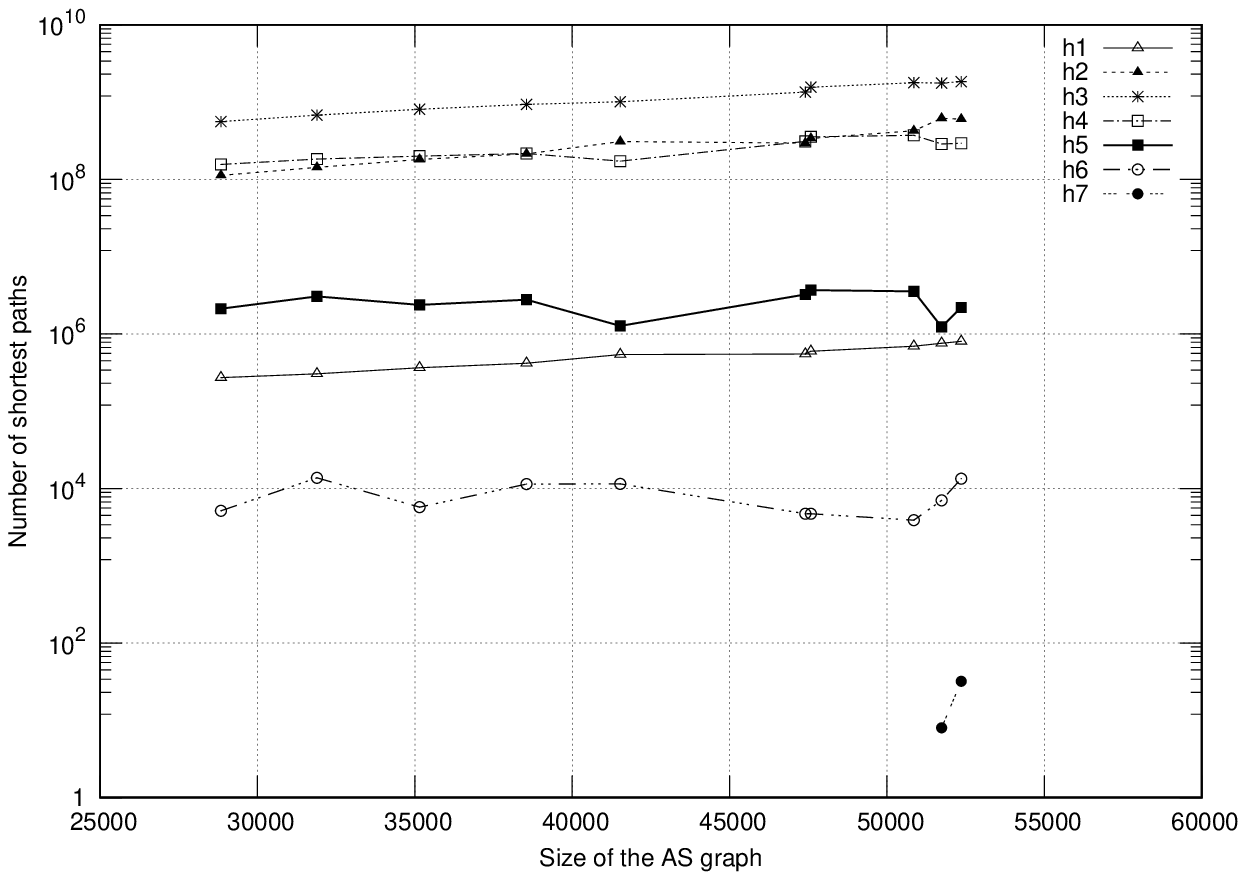}
\end{subfigure}
\caption{\textbf{Left}: shortest path distributions $s(h)$ for the AS graphs during the decade 2008-2017. \textbf{Right}: number of shortest paths of different lengths h$n$ as a function of the size of the AS graph. Here h$n$ indicates a shortest path whose length is $n$ hops.}
\label{fig_shortest_paths_distr_and_hops_vs_n}
\end{figure}

\subsection{Closeness centrality}\label{subsec:res_closeness_centr}
The closeness centrality $\Gamma$ as a function of the node's degree $k$ is shown in Fig.~\ref{fig_avgclosenesscentr_vs_degree_distr_2008_2017} for the years 2008 and 2017. The plots of the other years have similar slope. Their curves lie in between of those plotted  and are not shown in the figure for better readability because they overlap in the region $100 < k < 1000$. We observe that $\Gamma$ increases with the degree which means that big hub ASes are in the center of the Internet while low degree ASes are peripheric. We consider $\Gamma$ in three regions: $k \le 100$, $100 < k \le 1000$ and $k > 1000$ corresponding respectively to low, medium and high degree and we find that within errors it is almost constant over the period 2008-2017 with average  values of $0.392 \pm 0.007$, $0.434 \pm 0.004$ and $0.484 \pm 0.004$.  

\begin{figure}[htpb]
  \begin{center}
    \includegraphics[width=3in]{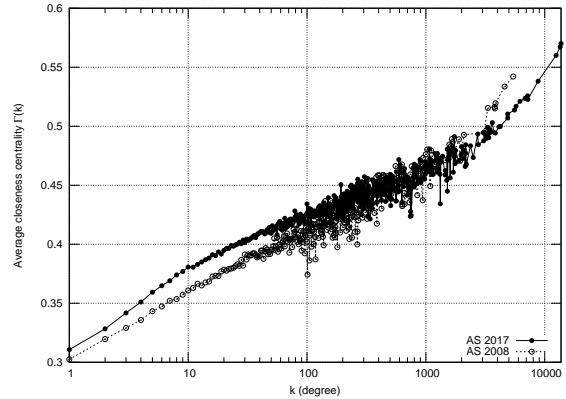}
\caption{Average closeness centrality $\Gamma(k)$ as a function of the degree $k$ for the AS graphs of the years 2008 and 2017.}
\label{fig_avgclosenesscentr_vs_degree_distr_2008_2017}
  \end{center}
\end{figure}

\subsection{Betweenness centrality}\label{subsec:res_betw_centr}
The average node betweenness centrality as a function of the degree $k$ is shown in Fig.~\ref{fig_avgnodebetwcentr_vs_degree_distr_2008_2017} for the AS graphs of the years 2008 and 2017. As in the case of the closeness centrality, we do not plot the curves of the other years for readability reasons. However, for all the years the average node betweenness centrality increases with the degree which means that the higher is the degree of an AS the more is the number of shortest paths passing through it. There is evidence of an overall slight decrease of $B_n(k)$ during the evolution of the Internet from 2008 to 2017. The overall average values of $B_n$ calculated in 2008 and 2017 are $\sim$7.1$\times 10^{-5}$ and $\sim$3.6$\times 10^{-5}$ respectively. In Tab.~\ref{tab_avgbetwcentr} are reported the average values of $B_n(k)$ calculated in the degree regions $k \leq 100$, $100 < k \leq 1000$ and $ k > 1000$. 

In order to study $B_e$ we represent an edge as a point of the $xy$ plane whose coordinates ($k_x$, $k_y$) are the degrees of the nodes it connects. In Fig.~\ref{fig_avgedgebetwcentr_k_k} is shown, for each year of the decade 2008-2017, the colored 3D map of the average $B_e$. The highest $B_e$ is associated to edges which have at least a high degree ($k > 1000$) AS as a terminal. Edges connecting low or medium degree ASes have lower $B_e$. This is what one would expect considering that high degree ASes are the backbone of the Internet and the most part of the shortest routes should cross them. We also observe a  slight decrease of $B_e$ over time. The overall average $B_e$ was $\sim$2.2$\times 10^{-5}$ in 2008 and $\sim$0.7$\times 10^{-5}$ in 2017, indicating that somehow the Internet has become less congested although it has expanded. By looking at the colored contour maps shown on the right side of Fig.~\ref{fig_avgedgebetwcentr_k_k} we infer that during its evolution the lowering of $B_e$ affected first the part of the Internet containing low and medium degree ASes ($k < 1000$) and subsequently the backbone. The overall average values of $B_n$ and $B_e$ were measured also in~\cite{Mahadevan_2006} for three sources of data. Authors found that for the AS graph of the Internet constructed from the CAIDA Skitter~\cite{Skitter} repository with data collected in March 2004 $B_n$ and $B_e$ were $\sim$11.0$\times 10^{-5}$ and $\sim$5.4$\times 10^{-5}$. This is a further confirmation that during the evolution of the Internet the traffic load somehow decreases. This may be due to the adoption of more efficient routing policies and to infrastructural upgrades with  more advanced network devices.

\begin{figure}[htpb]
  \begin{center}
    \includegraphics[width=3in]{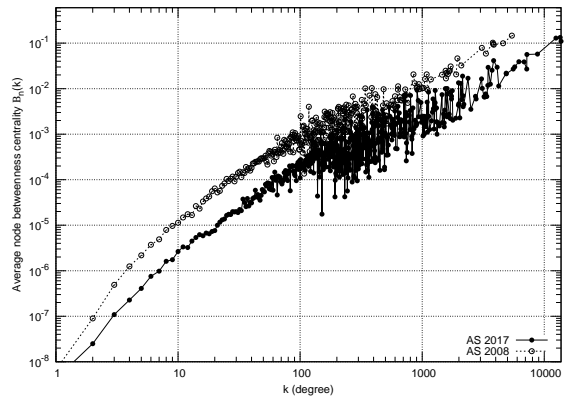}
\caption{Average node betweenness centrality $B_n(k)$ as a function of the degree $k$ for the AS graphs of the years 2008 and 2017.}
\label{fig_avgnodebetwcentr_vs_degree_distr_2008_2017}
  \end{center}
\end{figure}

\begin{table*}[htbp]
\begin{center}
\begin{adjustbox}{max width=\textwidth}
\begin{tabular}{|c||c|c|c|c|c|c|c|c|c|c|}
\hline
$B_n(k)$            & 2008   & 2009  & 2010   & 2011   & 2012    & 2013     & 2014  & 2015  & 2016  & 2017 \\
\hline
$k \leq 100$        & $3.3 \times 10^{-4}$  & $3.0 \times 10^{-4}$ & $2.2 \times 10^{-4}$  & $1.8 \times 10^{-4}$  & $1.2 \times 10^{-4}$   & $1.2 \times 10^{-4}$    & $1.2 \times 10^{-4}$ & $8.9 \times 10^{-5}$ & $7.7 \times 10^{-5}$ & $7.3 \times 10^{-5}$ \\
\hline
$100 < k \leq 1000$       & $3.0 \times 10^{-3}$ & $2.8 \times 10^{-3}$ & $2.2 \times 10^{-3}$ & $2.0 \times 10^{-3}$ & $1.5 \times 10^{-3}$ & $1.6 \times 10^{-3}$ & $1.5 \times 10^{-3}$ & $1.3 \times 10^{-3}$ & $1.0 \times 10^{-3}$ & $0.9 \times 10^{-3}$ \\
\hline
$ k > 1000$       & $4.5 \times 10^{-2}$ & $4.0 \times 10^{-2}$ & $3.1 \times 10^{-2}$ & $2.9 \times 10^{-2}$ & $1.9 \times 10^{-2}$ & $2.0 \times 10^{-2}$ & $1.8 \times 10^{-2}$ & $1.5 \times 10^{-2}$ & $1.4 \times 10^{-2}$ & $1.4 \times 10^{-2}$ \\
\hline
\end{tabular}
\end{adjustbox}
\end{center}
\caption{Average node betweenness centrality $B_n$ in the degree regions $k \leq 100$, $100 < k \leq 1000$ and $ k > 1000$.}
\label{tab_avgbetwcentr}
\end{table*}

\begin{figure}
  \begin{subfigure}[b]{0.4\textwidth}
    \includegraphics[width=\textwidth]{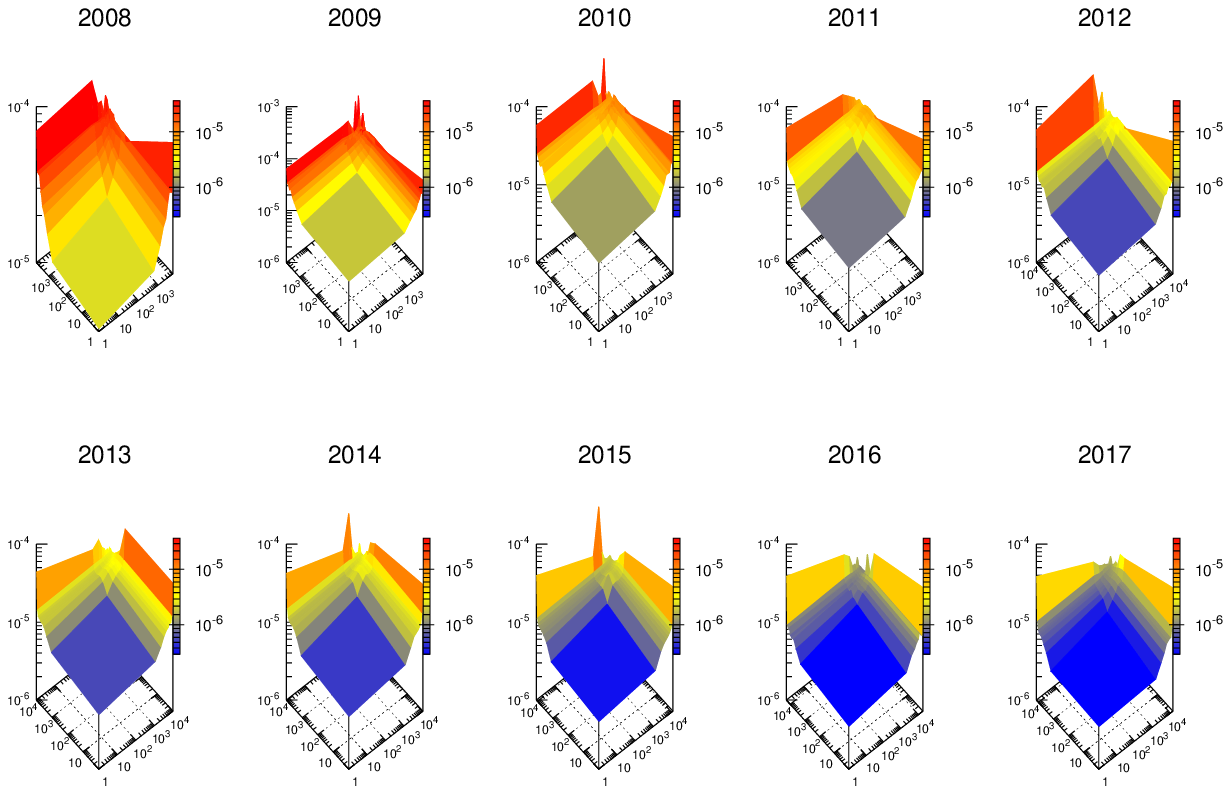}
    \label{fig_avgedgebetwcentr_k_k_all}
  \end{subfigure}
  \begin{subfigure}[b]{0.4\textwidth}
    \includegraphics[width=\textwidth]{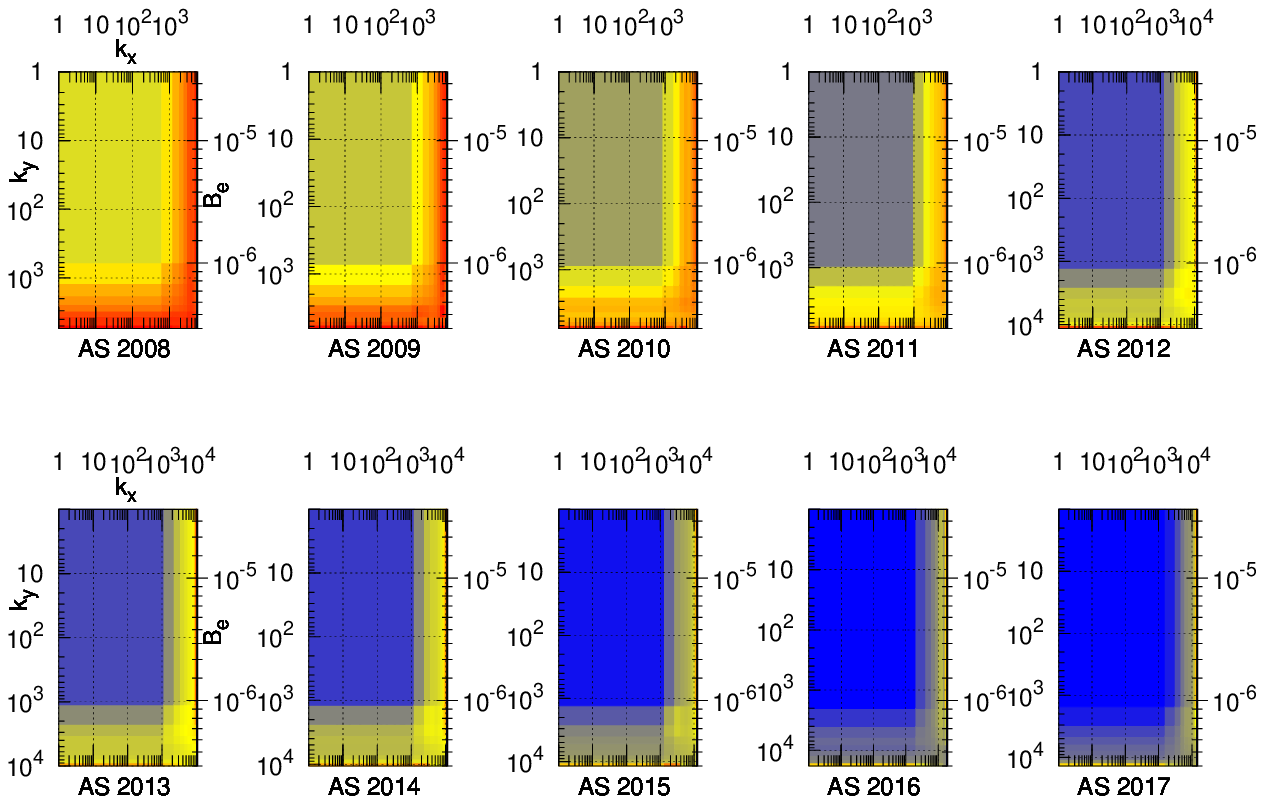}
    \label{fig_avgedgebetwcentr_k_k_all_map}
  \end{subfigure}
\caption{Average edge betweenness centrality $B_{e}$ for the AS graphs of the years 2008-2017. \textbf{Left}: color mapped 3D plots of $B_e$. To each edge is associated a point of the $xy$ plane whose coordinates ($k_x$, $k_y$) are the degree of the nodes it connects. \textbf{Right}: colored contour maps of the figures on the left.}
\label{fig_avgedgebetwcentr_k_k}
\end{figure}

\section{Conclusion}\label{sec:conclusion}
We studied the evolution of the Internet at the AS level during the decade 2008-2017. For each year of the decade we considered a snapshot of the AS undirected graph and analyzed how  a wide range of metrics related to structure, connectivity and centrality varies over time. During the decade 2008-2017 the Internet almost doubled its size and became more connected. The Internet is a scale free network because it contains both very high and low degree ASes. 
For all the years 2008-2017 the best fit of the degree distributions with a power law $P(k) \sim k^{-\alpha}$ provides values of the exponent very close to each other and around $\alpha \simeq 2.1$. However, the statistical analysis shows that a pure power law model fails to explain the scale free properties.
The study of the k-core decomposition shows that the Internet has a small internal nucleus composed of high degree ASes much more stable and connected than external cores. 
We investigated the hierarchical organization of the Internet by studying the average clustering coefficient $C$. We found that there are indications of an overall hierarchical organization of the Internet where a small fraction of big ASes are connected to many regions with high internal cohesiveness containing low and medium degree ASes and these regions are slightly connected among them. 
The average shortest path length $S$ of the Internet slightly decreased during the decade 2008-2017 form $\sim$3.1 to $\sim$2.9 measured in 2008 and 2016-2017 respectively. 
Regardless of the analyzed year, the closeness centrality $\Gamma$ of an AS increases with its degree. Hence, big ASes are in the center of the Internet and low degree ASes are in the periphery.
It is reasonable to assume that the traffic load of an AS or carried by an edge is proportional to the number of shortest paths passing through the AS and containing the edge. These measurements can be quantified by the average node and edge betweenness centrality $B_n$ and $B_e$. There is evidence of an overall slight decrease of both $B_n$ and $B_e$ during the decade 2008-2017, suggesting that during its evolution the Internet became less congested.

{\color{myaqua}

 \vskip 6mm

 \noindent\Large\bf Acknowledgments}

 \vskip 3mm

{ \fontfamily{times}\selectfont
 \noindent
The computing resources and the related technical support used for this work have been provided by CRESCO/ENEAGR\-ID High Performance Computing infrastructure and its staff~\cite{eneagrid}. CRESCO/ENEAGRID High Performance Computing infrastructure is funded by ENEA, the Italian National Agency for New Technologies, Energy and Sustainable Economic Development and by Italian and European research programmes, see https://www.eneagrid.enea.it for information.

 {\color{myaqua}

}}

\end{document}